\documentclass[11pt,notitlepage]{isasrep}
\usepackage[dvips]{graphicx}
\usepackage[small,hang]{caption2}
\begin{document} 

%%% Title of your paper %%%	                                           

\title{MID-IR Surveys - The Next Generations}
				
%%% Running title of your paper %%%

\markright{MID-IR Surveys - The Next Generations}

%%% Authors %%%

\author{
     Chris {\sc Pearson}\footnotemark[1]$\;$ 
       } 

%%% Date of Submission %%%

\date{(7 June 2000)} 

\maketitle	%% DO NOT CHANGE this line %%

%%% INPUT your affilations as the footnotes %%%

\renewcommand{\thefootnote}{\fnsymbol{footnote}} %% DO NOT CHANGE this line %%

\footnotetext[1]{Chris Pearson\\
The Institute of Space and Astronautical Science, 
3-1-1 Yoshinodai, Sagamihara, Kanagawa, 229-8510 Japan \\
E-mail: cpp@astro.isas.ac.jp }

\renewcommand{\thefootnote}{\arabic{footnote}}   %% DO NOT CHANGE this line %%
\setcounter{footnote}{0}			 %% DO NOT CHANGE this line %%
\renewcommand{\baselinestretch}{1}		 %% DO NOT CHANGE this line %%

%%% INPUT abstract here %%%

\begin{abstract}
Using new cosmological models for galaxy evolution, predictions for the next generation of deep mid-IR surveys are presented for both the ASTRO-F space telescope (due for launch 2003) and the proposed HII/L2 mission (2010). Although ASTRO-F will be able to detect between 20,000-40,000 sources in the mid-IR, deep surveys will be severely constrained by confusion due to faint sources (due to the relatively small aperture of all IR space telescopes to date). However, HII/L2 with its larger 3.5m mirror will be able to observe to fluxes 10-100 times deeper below the present confusion limits.
\end{abstract}

%%% Main body of the paper %%%

\section{ASTRO-F AND HII/L2}\label{sec:telescope}

\subsection{ASTRO-F}\label{sec:ASTRO-F}

ASTRO-F also known as the Infrared Imaging Surveyor (IRIS) is the second infrared astronomy mission of the Japanese Institute of Space and Astronautical Science (ISAS). ASTRO-F is a 70cm cooled telescope and will be  dedicated for infrared sky surveys (\cite{ona98}). The ASTRO-F Richey-Chretian telescope has a 70cm aperture and is cooled to 6K, and the detectors to 1.8K, using a light weight liquid Helium Cryostat. Two 2-stage Stirling-cycle coolers ensure minimum heat flow from the outer wall of the cryostat, almost doubling the lifetime of the Helium thus allowing approximately 150l of liquid  Helium to sustain the telescope for more than 400 days. Another advantage of using the mechanical coolers is that the near-infrared detectors will still be usable even after the Helium expires (providing the coolers continue to function). 

ASTRO-F covers wide wavelength range from the K-band to
200$\mu$m. Two focal-plane instruments are installed.  The first is the Far-Infrared Surveyor (FIS, \cite{kaw98} ) which will survey the entire sky in the wavelength range from 50 to 200$\mu$m (\cite{take99}, \cite{cpp001}). The other focal-plane instrument is the Infrared Camera (IRC, see Table~\ref{IRC}). It employs large-format detector arrays and will take deep images of selected sky regions in the near and mid infrared range (\cite{cpp002}).

 The IRC is a wide-field imaging instrument (\cite{mat98}) consisting of three independent camera systems. IRC-NIR (1.8-5 $\mu$m), MIR-S (5-12$\mu$m) and MIR-L (10-25$\mu$m) - see Table~\ref{IRC}  (\cite{wat00}), capable of simultaneously observing 3 different fields (10'x10' FOV each) of the sky separated by approximately 20\arcmin, with a diffraction-limited spatial resolution of approximately 2\arcsec. At the aperture stop of each camera (i.e. the image position of the telescope primary mirror, $\approx$11$\sim$12mm in diameter), a 6-position filter wheel is placed to select the observing wavelength band. The filter bands are selected by rotating the filter wheels via commands.

 ASTRO-F will have a much higher sensitivity than that of the IRAS survey having 50-100 times higher sensitivity at 100$\mu$m and more than 1000 times that at mid-infrared wavelengths. Table~\ref{limits1} shows the current expected sensitivities for a single pointing ($5\sigma$) using the narrow band filters on the IRC-MIR detectors. With the ASTRO-F surveys, great progress is expected in the research on evolution of galaxies, formation of stars and planets, dark matter and brown dwarfs. ASTRO-F is now scheduled to be launched with ISAS's M-V launch vehicle, into a sun-synchronous polar orbit with an altitude of 750 km in mid 2003.

\begin{table}
\begin{center}
\caption{ASTRO-F Infra-Red Camera and HII/L2 detector Parameters}
\renewcommand{\arraystretch}{1.4}
\setlength\tabcolsep{15pt}
\begin{tabular}{@{}lp{6.0cm}l}
\hline\noalign{\smallskip}
Channel & Wavelength Bands & FOV (Pixel size) \\
\noalign{\smallskip}
\hline
\noalign{\smallskip}
IRC-NIR & K,L,M+($1.25-2.5, 2-5.5\mu$m grism) & $10'$x$10'(1.4''/pixel)$ \\
IRC-MIR-S & $7,9,11\mu$m($5-10\mu$m grism) & $10'$x$10'(2.34''/pixel)$ \\
IRC-MIR-L & $15,20,25\mu$m($10-25\mu$m grism) & $10'$x$10'(2.34''/pixel)$ \\
\noalign{\smallskip}
\hline
\noalign{\smallskip}
SW-MIR & $5-12\mu$m & $6.1'$x$6.1'(0.18''/pixel)$ \\
LW-MIR & $12-25\mu$m & $6.1'$x$6.1'(0.36''/pixel)$ \\
\noalign{\smallskip}
\hline
\noalign{\smallskip}
\end{tabular}\\
\label{IRC}
\end{center}
\end{table}

\subsection{HII/L2}\label{sec:HII/L2}
 
Although the current generation of IR astronomical satellites ISO (\cite{kessler96}), SIRTF (\cite{rieke00}) and ASTRO-F have produced or will produce excellent results they all suffer from at least some of the following constraints. The mission lifetime is relatively short, limited by the capacity of liquid Helium coolant. The satellites have to avoid radiation from the Sun and Earth thus the single integration times that are capable are relatively short (10mins. for ASTRO-F). Finally, the actual apertures of the telescopes are relatively small (70cm for ASTRO-F, 85cm for SIRTF) due to the weight constraints placed on the mission payloads by the cooling cryostats. These small aperture telescopes will undoubtedly suffer from severe constraints in sensitivity due to source confusion due to faint sources (e.g. \cite{serjeant97}, \cite{cpp002}).

To overcome these constraints, the HII/L2 mission has been proposed. The  HII/L2 mission is a next generation mission proposed by the Japanese Institute of Space and Astronautical Science (ISAS) tentatively scheduled for launch in 2010 from the new H-IIA launch vehicle  (\cite{naka98}). The HII/L2 would be a {\it warm launch\ } cooled telescope, launched at ambient temperature and cooled to 4.5K in space via natural radiation cooling and a cryogenic cooler. The lack of a large cryostat means that the satellite weight is vastly reduced and significantly larger aperture mirror can be accommodated. In the case of HII/L2 a 3.5m mirror (or 8m unfoldable). In addition, the HII/L2 telescope would be put into a {\it halo} orbit around the 2nd Sun-Earth Lagrangian liberation point (S-E L2). At this distance, the apparent size of the Earth is reduced to $\sim$30\arcmin , thereby greatly reducing the heat from the Earth. Furthermore, the Earth, Sun and Moon are approximately in the same direction enabling easier shielding and longer integration times of larger areas of the sky.  

In the mid-infrared region HII/L2 would envisage having 2 main instruments covering a short wavelength range from 5-12$\mu$ and longer wavelengths from 12-2$\mu$ with a FOV of 6.1x6.1$\arcmin$ (see tab.~\ref{IRC}). The short wavelength detector would consist of a 2x2 array each of 1024x1024 pixels. The longer wavelength detector would be a single 1024x1024 pixel array although the total size of the observable area would be 368.64x368.64sq.arcsec $\approx$ 0.01 sq.deg..

HII/L2 would cover the wavelength range from 5-200$\mu$m, thus complementing both NGST (\cite{mather00}) and FIRST (\cite{pilbratt00}) which excel at shorter and longer wavelengths respectively. Expected sensitivities of HII/L2 are shown in tab.~\ref{limits1}

\section{MID-INFRARED MODEL PARAMETERS}\label{sec:model}

Galaxy source counts are simulated by using an extended and improved evolution of the Pearson and Rowan-Robinson model (\cite{cpp96}). The model utilizes of a 4 component parameterization consisting of normal, starburst, ultraluminous infrared galaxies \& an AGN (Seyfert/QSO 3-30$\mu$m dust torii) component (\cite{RR95}). Components are distinguished on a basis of SED and luminosity class. The composite cool and warm 60$\mu$m luminosity functions of \cite{saun90} are used to represent the normal, starburst and ULIG galaxies respectively. The AGN luminosity function is defined at 12$\mu$m (\cite{rush93}). K-corrections are calculated using model Spectral Energy Distribution (SED's) templates for each galaxy population (\cite{esf001}, \cite{esf002}, \cite{RR95}). The new galaxy SED templates also incorporate the full range of PAH features crucial for mid-infrared predications. Both luminosity evolution and density evolution is incorporated into the models allowing the source counts from sub-mm to NIR wavelengths to be fitted by one consistent model (see \cite{cpp003} for detailed description of these new models).

\begin{figure}[ht] 
   \begin{center}
   \includegraphics[angle=0,width=16cm]{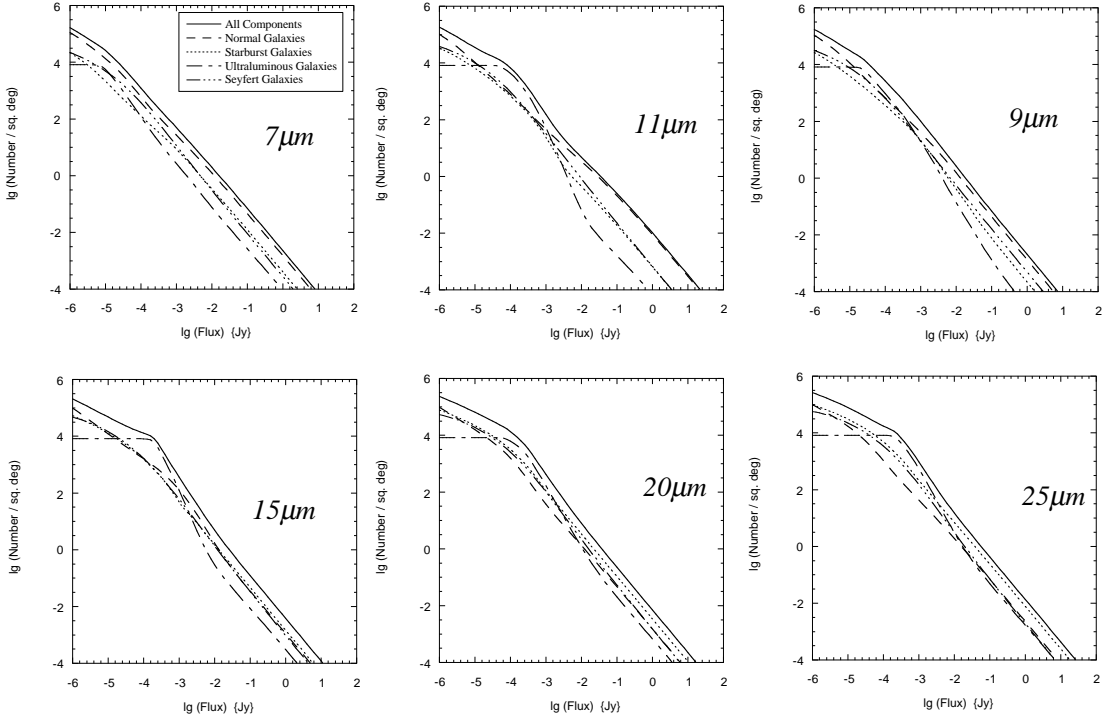}
   \caption{Integral galaxy source counts, as described in the text, at 7, 9, 11, 15, 20 \& 25$\mu$m respectively, corresponding to the IRC-MIR-S and IRC-MIR-L filter wavelengths. Total source counts are shown ({\it solid line\,}) with normal ({\it dash\,}), starburst ({\it dot}), ultraluminous galaxy ({\it dash dot\,}) and AGN components ({\it dot-dot-dot-dash\,}).}
   \label{intcts}
   \end{center}
\end{figure}

\section{SURVEY STRATEGIES AND DETECTION LIMITS}\label{sec:survey}

\subsection{Deep mid-IR surveys with ASTRO-F \& HII/L2}\label{sec:strategy}

For ASTRO-F an ultra-deep survey covering $\approx$ 3200 sq. arcmin. around the North Ecliptic Pole (NEP) is investigated. The 3200sq.arcmin. survey would take an extremely deep (20 pointings) observation around a $\sim$67\arcmin  diameter doughnut of width $\sim$10\arcmin  (i.e. the field of view of the IRC) around the NEP (\cite{cpp002}). The aim of this survey would be to image as deep as possible in all bands of the MIR-IRC. 

For the case of HII/L2, an extremely preliminary survey strategy is investigated. A single 1 hour pointing covering the FOV of 0.01sq.deg., using the short and long wavelength detectors. For the sake of comparison, equivalent wavelengths to the ASTRO-F IRC-MIR-S \& MIR-L channels are selected (7, 9, 11, 15, 20, 25$\mu$m).

Fig.~\ref{intcts} shows the integral counts at these mid-IR wavelengths assuming the model described in Sec.~\ref{sec:model}. In general, the normal galaxies dominate at the shortest wavelengths. Towards longer wavelengths the dominance of the starburst and ULIG galaxies increase being approximately equal to the normal galaxies at 15$\mu$m and then surpassing them in the 20-25$\mu$m bands. The {\it bump} in the ULIG counts at faint fluxes is caused by the strong density evolution incorporated into these models.

Table~\ref{limits1} shows the current expected sensitivities for a deep 20 pointing ($5\sigma$) survey using the narrow band filter wavelengths on the ASTRO-F IRC-MIR detectors for ASTRO-F and a single 1 hour deep pointing on HII/L2. Using the predictions made by the cosmological source counts the confusion limit due to background sources can be calculated and is tabulated in table~\ref{limits1}. For any one space telescope, the confusion limit at mid-infrared wavelengths will be lower than in the far-infrared (and lie at higher redshift) due to the higher spatial resolution at the shorter wavelengths ($ \sim \lambda /D$).  The confusion limits are calculated assuming the classical confusion criteria of a source density of 1 source per 40 beams of the observing instrument, where the beam diameter is given by $d=1.2 \lambda /D$, where $D$ is the telescope diameter (70cm \& 3.5m for ASTRO-F and HII/L2 respectively).

\begin{table}[htbp]
\begin{center}
\caption{ASTRO-F \& HII/L2 Survey Detection \& Confusion Limits.}
\label{limits1}
\begin{tabular}{@{}lcccc}
wavelength & \multicolumn{4}{c}{Sensitivity limit (micro Jy)} \\
   &  \multicolumn{2}{c} {ASTRO-F} &  \multicolumn{2}{c} {HII/L2} \\
   & Detector & Confusion &  Detector & Confusion   \\
\hline
$7\mu$m & $4$ & $3.3$ & $0.163$ & $0.001$ \\
$9\mu$m & $7$ & $7.8$ & $0.323$ & $0.014$ \\
$11\mu$m & $12$ & $14.5$ & $0.528$ & $0.066$ \\
$15\mu$m & $19$ & $76$ & $0.926$ & $0.355$ \\
$20\mu$m & $27$ & $158$ & $1.323$ & $1.259$ \\
$25\mu$m & $34$ & $363$ & $1.633$ & $3.981$ \\
\hline
\end{tabular}
\end{center}
\end{table}

\begin{figure}[ht] 
   \begin{center}
   \includegraphics[angle=0,width=16cm]{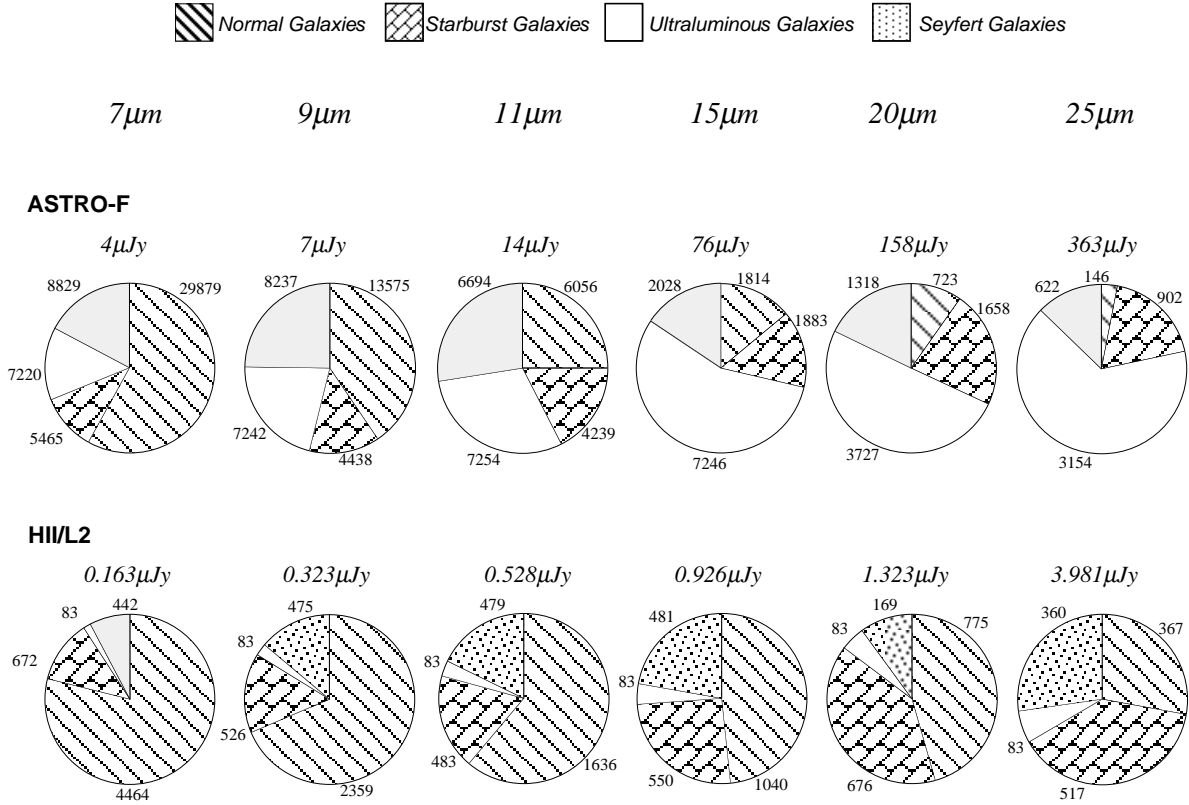}
   \caption{Summary of ASTRO-F deep 3200sq.arcmin. ({\it top}) and HII/L2 0.01sq.deg. ({\it bottom}) Mid-IR surveys. Predicted numbers and relative proportions of galaxies are shown for the respective survey sensitivity limits (in $\mu Jy$) in table~\ref{limits1}. Note that if the survey sensitivity is source confusion limited then the sensitivity limit is dependent on the evolutionary model.}
   \label{pie}
   \end{center}
\end{figure}

\subsection{Survey Predictions}\label{sec:predict}

In Fig.~\ref{pie} the predictions for deep surveys with both ASTRO-F and HII/L2 are shown.

\begin{figure}[ht] 
   \begin{center}
   \includegraphics[angle=0,width=16cm]{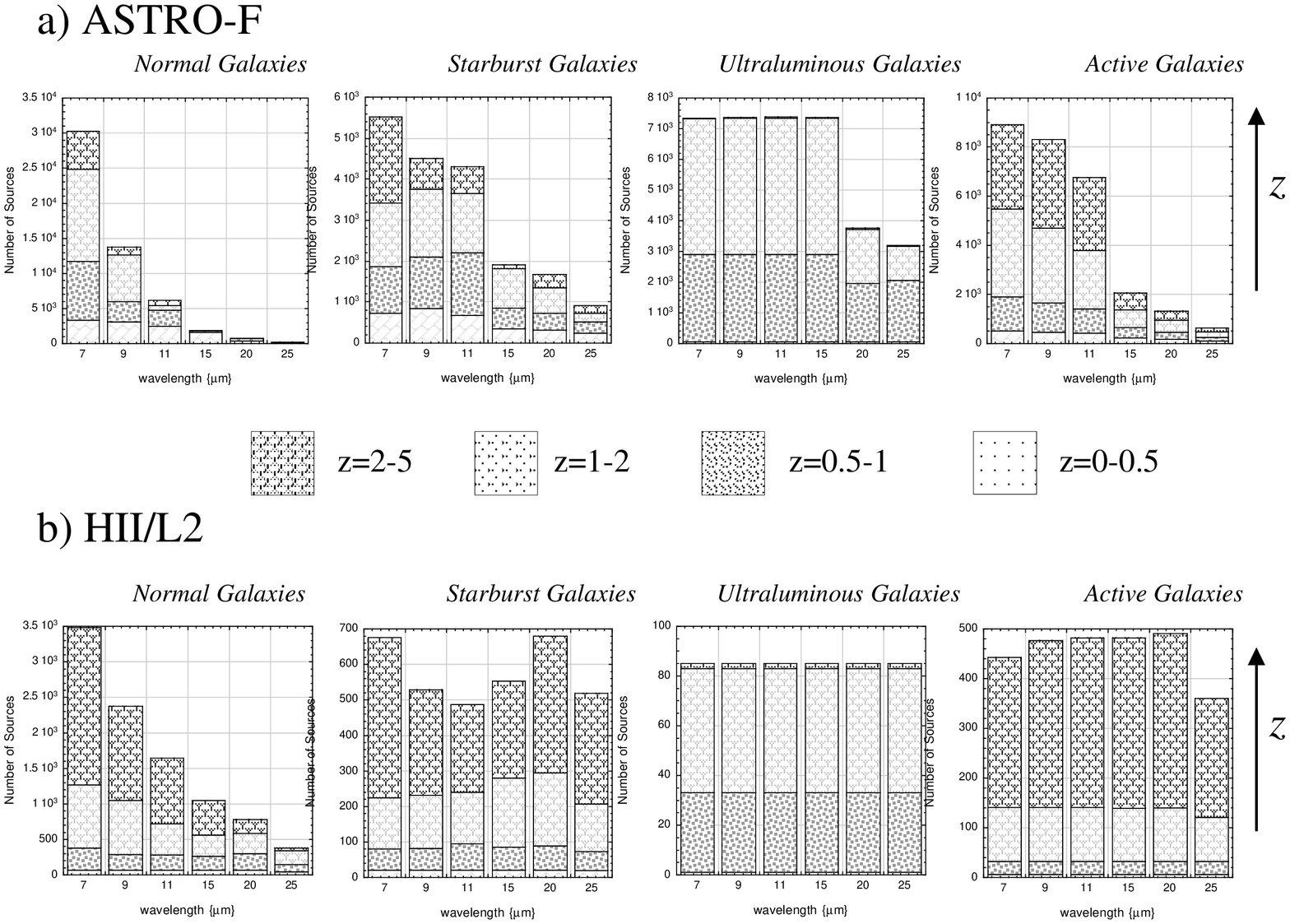}
   \caption{N-z distributions for ASTRO-F deep 3200sq.arcmin. ({\it top}) and HII/L2 0.01sq.deg. ({\it bottom}) mid-IR surveys. Predicted numbers and relative proportions of galaxies are shown for the respective survey sensitivity limits (in $mJy$) in table~\ref{limits1}. Note the scales on the axes differ}
   \label{graph}
   \end{center}
\end{figure}

The predictions for the proposed 3200sq.arcmin. deep NEP MIR survey with ASTRO-F are shown as a function of object class (normal, starburst, ULIG and AGN). Of the order of 20,000-40,000 sources would be detected with the IRC-MIR-S instrument (7-11$\mu$m), the majority of which would be normal galaxies that have relatively strong K-corrections as the near-infrared/optical emission is sampled out to higher redshifts (the 3.3$\mu$m UIB feature being sampled at z$\sim$1.1 and the K-band at a redshift of $\sim$ 2.2 for the 7$\mu$m band). Of the order of 4000 starburst galaxies and 8000 AGN would be seen in the short wavelength bands decreasing towards the longer (20-25$\mu$m) wavelength bands where the confusion limit becomes severe. In general the trend is towards an increasing fraction of starburst galaxies to normal galaxies towards longer wavelengths. The AGN component maintains an almost constant fraction of the total observed sources over the entire 7-25$\mu$m wavelength range. The fraction of ULIGs rises from around 10$\%$ in the shortest waveband to 65$\%$ in the longest. The number of ULIGs predicted remains approximately constant $\sim$7000 from 7-15$\mu$m, dropping to $\sim$3000 at the longest wavelengths due in part to the relatively lower sensitivity of these bands. In the 7-15$\mu$m bands we are essentially seeing the entire ULIG population.  A significant contribution comes from the PAH features in the ULIG SED at the longer wavelengths. At all wavebands the contribution from the ULIG component peaks at about redshift 1 due to the form of the strong density evolution assumed by the model. Also prominent at 20$\mu$m is the 9.7$\mu$m silicate absorption line in the starburst and ULIG SEDs, producing a prominent double hump in the ULIG number redshift distribution. This feature is prominent due to the evolution incorporated into the ULIG model. The emission from the 3.3$\mu$m feature enhances the detectability of the normal galaxy component (and to a similar extent the starburst component) out to redshifts of $\sim$1-2.5 from 7-11$\mu$m respectively (see fig.~\ref{graph}). In fact as much as 60$\%$ of the normal galaxies may lie at z$>$1 in the shortest 2 bands. In all bands approximately 50$\%$ of the starburst galaxies lie at z$>$1 within which $\approx$10-20$\%$ may lie at z$>$2. Interestingly, the role of the mid-IR PAH features seems rather important enhancing the starburst galaxy population at redshifts of approximately 1 and 2 in the 15 and 20-25$\mu$m bands respectively as the mid-infrared region containing the 6-13$\mu$m {\it forest} of PAH features is redshifted into the respective observation windows.
For HII/L2, the survey predictions are very different because the nature of the survey is essentially a {\it narrow beam\ } penetrating to significantly high redshifts and lower flux densities. In this scenario the proportion of ULIG sources is relatively low due to the fact that they are most numerous at redshifts of $\approx$1.The dominant population at short wavelengths are the normal galaxies (over 75$\%$). A large contribution of the flux at these wavelengths comes from the 3.3$\mu$m UIB feature. At the longest wavelengths the starburst galaxies become the dominant population. The difference between the 2 surveys (ASTRO-F \& HII/L2) can also be readily seen from the N-z distributions in fig.~\ref{graph}. The proportion of high redshift galaxies is significantly higher in the HII/L2 survey than the ASTRO-F survey. In fact, in the case of the ULIG component we are essentially seeing the entire MIR population of these objects.

\section{SUMMARY \& CONCLUSIONS}\label{sec:summary}

Two possible survey strategies utilizing ASTRO-F (the Imaging Infra Red Surveyor - IRIS) and the proposed HII/L2 mission have been investigated.

For ASTRO-F an ultra-deep survey covering $\sim 3200$sq.arcmin. in a doughnut around the north ecliptic pole was investigated. For HII/L2 a deep, narrow beam 0.01sq.deg. survey was investigated assuming a 1 hour pointing. Survey predictions have been made assuming evolutionary scenarios including normal, evolving starburst \& AGN galaxy components and a strongly evolving ULIG component (\cite{cpp003}). 
For the ASTRO-F survey, in general the models predict that between 20,000-40,000 sources may be detected in the shortest wavebands dropping to  $\approx$5000 in the longest (25$\mu$m) band. At the longest MIR wavelengths, the deep survey would be severely constrained by the source confusion limit rendering the deepest integrations relatively futile. This will be an important constraint for both the ASTRO-F, SIRTF and future relatively small aperture space infrared telescopes. Although SIRTF has a slightly larger primary (85cm) than ASTRO-F, source confusion at longer mid-infrared wavelengths will still be severe.  To integrate deeper {\it beneath} the current constraints set by the source confusion, telescopes with significantly larger mirrors will be required (e.g. NGST \& HII/L2 mission). In the shorter bands it should be possible to detect many sources out to high redshift $\sim 5$ with more than half of the normal galaxies being at redshift $>$1 in the 7 \& 9$\mu$m bands. Without doubt the infrared unidentified bands (UIB - PAH features) aid in the detection of galaxies to higher redshifts, with the 3.3$\mu$m enhancing the detectabilty of normal galaxies out to high redshift and the PAH {\it forest} in general enhancing the detectability of all starforming and normal galaxies in the 20$\mu$m band.

HII/L2, with its significantly larger aperture (and therefore resolution) will be able to observe below the confusion limit that constrains ASTRO-F to fluxes between 10-100 times deeper. A 0.01sq.deg. survey would expect to see $\>$1000 sources ($\sim$500 starbursts, $\sim$400 AGN) out to high redshift. An interesting possibility for HII/L2 would be the inclusion of an almost monochromatic filter to trace the unidentified IR bands in the PAH {\it forest} out to high redshift (or in fact as a direct indicator of redshift?)

As it is hoped that ASTRO-F and SIRTF will complement each other and pave the way for future generations of space telescopes, so it is also enthusiastically hoped that HII/L2 may would play a complementary role to NGST, FIRST and Planck.

\section*{ACKNOWLEDGMENTS}

 CPP is supported by a Japan Society for the Promotion of Science (JSPS) fellowship and is grateful for all the members of the ASTRO-F project for their efforts and support during his time in Japan at ISAS.

%%% 	REFERENCES (do not forget {}) %%%

\end{document}